# Cognitive computational neuroscience

**Nikolaus Kriegeskorte**[a] & **Pamela K. Douglas**[b]

   a. Department of Psychology, Department of Neuroscience, Department of Electrical Engineering, Zuckerman Mind Brain Behavior Institute, Columbia University, New York, NY
   b. Center for Cognitive Neuroscience, University of California, Los Angeles, CA

**To learn how cognition is implemented in the brain, we must build computational models that can perform cognitive tasks, and test such models with brain and behavioral experiments. Cognitive science has developed computational models of human cognition, decomposing task performance into computational components. However, its algorithms still fall short of human intelligence and are not grounded in neurobiology. Computational neuroscience has investigated how interacting neurons can implement component functions of brain computation. However, it has yet to explain how those components interact to explain human cognition and behavior. Modern technologies enable us to measure and manipulate brain activity in unprecedentedly rich ways in animals and humans. However, experiments will yield theoretical insight only when employed to test brain-computational models. It is time to assemble the pieces of the puzzle of brain computation. Here we review recent work in the intersection of cognitive science, computational neuroscience, and artificial intelligence. Computational models that mimic brain information processing during perceptual, cognitive, and control tasks are beginning to be developed and tested with brain and behavioral data.**

Understanding brain information processing requires that we build computational models that are capable of performing cognitive tasks. The argument in favor of task-performing computational models was well articulated by Allen Newell in his commentary "You can't play 20 questions with nature and win" in 1973.[1] Newell was criticizing the state of cognitive psychology. The field was in the habit of testing one hypothesis about cognition at a time, in the hope that forcing nature to answer a series of binary questions would eventually reveal the brain's algorithms. Newell argued that testing verbally defined hypotheses about cognition might never lead to a computational understanding. Hypothesis testing, in his view, needed to be complemented by the construction of comprehensive task-performing computational models. Only synthesis in a computer simulation can reveal what the interaction of the proposed component mechanisms actually entails, and whether it can account for the cognitive function in question. If we did have a full understanding of an information-processing mechanism, then we should be able to engineer it. "What I cannot create, I do not understand" in the words of physicist Richard Feynman, who left this sentence on his blackboard when he died in 1988.

Here we argue that task-performing computational models that explain how cognition arises from neurobiologically plausible dynamic components will be central to a new cognitive computational neuroscience. We first briefly trace the steps of the cognitive and brain sciences and then review several exciting recent developments that suggest that it might be possible to meet the combined ambitions of cognitive science (to explain how humans learn and think)[2] and computational neuroscience (to explain how brains adapt and compute)[3] using neurobiologically plausible artificial intelligence (AI) models.

In the spirit of Newell's critique, the transition from cognitive psychology to cognitive science was defined by the introduction of task-performing computational models. Cognitive scientists knew that understanding cognition required AI and brought engineering to cognitive studies. In the 1980s, cognitive science achieved important advances with symbolic cognitive architectures[4, 5] and neural networks,[6] using human behavioral data to adjudicate between candidate computational models. However, computer hardware and machine learning were not sufficiently advanced to simulate cognitive processes in their full complexity. Moreover, these early developments relied on behavioral data alone and did not leverage constraints provided by the anatomy and activity of the brain.

With the advent of human functional brain imaging, scientists began to relate cognitive theories to the human brain. This endeavor started with electroencephalography (EEG),[7] expanded with magnetoencephalography (MEG)[8] and positron emission tomography (PET), and exploded with the invention of functional magnetic resonance imaging (fMRI).[9] It came to be called cognitive neuroscience.[10]

Cognitive neuroscientists began by mapping cognitive psychology's boxes (information-processing modules) and arrows (interactions between modules) onto the brain. This was a step forward in terms of engaging brain activity, but a step back in terms of computational rigor. Methods for testing the task-performing computational models of cognitive science with brain-activity data had not been conceived. As a result, cognitive science and cognitive neuroscience parted ways in the 1990s.

Cognitive psychology's tasks and theories of high-level functional modules provided a reasonable starting point for mapping the coarse-scale organization of the human brain with functional imaging techniques, including EEG, PET and early fMRI, which had low spatial resolution. Inspired by cognitive psychology's notion of module,[11] cognitive neuroscience developed its own game of 20 questions with nature. A given study would ask whether a particular cognitive module could be found in the brain. The field mapped an ever increasing array of cognitive functions to brain regions, providing a useful rough draft of the global functional layout of the human brain.

Brain mapping enables us to relate the performance of a task to activity all over the brain, using statistical inference techniques that account for the multiple testing across locations.[12] As imaging technology advances, increasingly detailed patterns of selectivity can be mapped across the brains of humans and animals. In humans, fMRI affords up to whole-brain coverage at resolutions on the order of a millimeter; in animals, modern techniques, such as calcium imaging, can capture vast numbers of neurons with single-neuron resolution.

A brain map, at whatever scale, does not reveal the computational mechanism (Figure 1). However, mapping does provide constraints for theory. After all, information exchange incurs costs that scale with the distance between the communicating regions – costs in terms of physical connections, energy, and signal latency. Component placement is likely to reflect these costs. We expect regions that need to interact at high bandwidth and short latency to be placed close together.[13] More generally, the topology and geometry of a biological neural network constrain its dynamics, and thus its functional mechanism. The literature on functional localization results, especially in combination with anatomical connectivity, may therefore ultimately prove useful for modeling brain information processing.

Modern meta-analysis techniques for brain imaging data enable us to go beyond localization of predefined cognitive components and learn about the way cognition is decomposed into component functions.[14] The field has also gone beyond associating overall activation of brain regions with their involvement in particular functions. A growing literature aims to reveal the representational content of brain regions by analyzing their multivariate patterns of activity.[15,16,17,18]

Despite methodological challenges,[19,20] many of the findings of cognitive neuroscience provide a solid basis to build on. For example, the findings of face-selective regions in the human ventral stream[21] have been thoroughly replicated and generalized.[22] Nonhuman primates probed with fMRI exhibited similar face-selective regions,[23] which had evaded explorations with invasive electrodes, because the latter do not provide continuous images over large fields of view. Localized with fMRI and probed with invasive electrode recordings, the primate face patches revealed high densities of face-selective neurons,[24] with invariances emerging at higher stages of hierarchical processing, including mirror-symmetric tuning and view-tolerant representations of individual faces in the anterior-most patch.[25] The example of face perception illustrates, on one hand, the solid progress in mapping the anatomical substrate and characterizing neuronal responses [26] and, on the other, the lack of definitive computational models. The literature does provide clues to the computational mechanism. A brain-computational model of face recognition[27] will have to explain the spatial clusters of face-selective units and the selectivities and invariances observed with fMRI[28,29] and invasive recordings.[25,30]

Cognitive neuroscience has mapped the global functional layout of the human and nonhuman primate brain.[31] However, it has not achieved a full computational account of brain information processing. The challenge ahead is to build computational models of brain information processing that are consistent with brain structure and function and perform complex cognitive tasks. The following recent developments in cognitive science, computational neuroscience, and artificial intelligence suggest that this may be achievable.

(1) Cognitive science has proceeded from the top down, decomposing complex cognitive processes into their computational components. Unencumbered by the need to make sense of brain data, it has developed task-performing computational models at the cognitive level. One success story is that of Bayesian cognitive models, which optimally combine prior knowledge about the world with sensory evidence.[32,33,34,35] Initially applied to basic sensory and motor processes,[35,36] Bayesian models have begun to engage complex cognition, including the way our minds model the physical and social world.[2] These developments occurred in interaction with statistics and machine learning, where a unified perspective on probabilistic empirical inference has emerged. This literature provides essential computational theory for understanding the brain. In addition, it provides algorithms for approximate inference on generative models that can grow in complexity with the available data – as might be required for real-world intelligence.[37,38,39]

(2) Computational neuroscience has taken a bottom-up approach, demonstrating how dynamic interactions between biological neurons can implement computational component functions. In the past two decades, the field developed mathematical models of elementary computational components and their implementation with biological neurons.[40,41] These include components for sensory coding,[42,43] normalization,[44] working memory,[45] evidence accumulation and decision mechanisms,[46,47,48] and motor control.[49] Most of these component functions are computationally simple, but they provide building blocks for cognition. Computational neuroscience has also begun to test complex computational models that can explain high-level sensory and cognitive brain representations.[50,51]

(3) Artificial intelligence has shown how component functions can be combined to create intelligent behavior. Early AI failed to live up to its promise, because the rich world knowledge required for feats of intelligence could not be either engineered or automatically learned. Recent advances in machine learning, boosted by growing computational power and larger data sets to learn from, have brought progress at perceptual,[52] cognitive,[53] and control challenges.[54] Many advances were driven by cognitive-level symbolic models. Some of the most important recent advances are driven by deep neural network models, composed of units that compute linear combinations of their inputs, followed by static nonlinearities.[55] These models employ only a small subset of the dynamic capabilities of biological neurons, abstracting from fundamental features such as action potentials. However, their functionality is inspired by brains and could be implemented with biological neurons.

The three disciplines contribute complementary elements to biologically plausible computational models that perform cognitive tasks and explain brain information processing and behavior (Figure 2). Here we review the first steps in the literature toward a cognitive computational neuroscience that meets the combined criteria for success of cognitive science (computational models that perform cognitive tasks and explain behavior) and computational neuroscience (neurobiologically plausible mechanistic models that explain brain activity). If computational models are to explain animal and human cognition, they will have to perform feats of intelligence. Machine learning and AI more broadly are therefore key disciplines that provide the theoretical and technological foundation for cognitive computational neuroscience.

The overarching challenge is to build solid bridges between theory (instantiated in task-performing computational models) and experiment (providing brain and behavioral data). The first part of this article describes bottom-up developments that begin with experimental data, and attempt to build bridges from the data in the direction of theory.[56] Given brain-activity data, connectivity models aim to reveal the large-scale dynamics of brain activation; decoding and encoding models aim to reveal the content and format of brain representations. The models employed in this literature provide constraints for computational theory, but they do not in general perform the cognitive tasks in question and, thus, fall short of explaining the computational mechanism underlying task performance.

The second part of this article describes developments that proceed in the opposite direction, building bridges from theory to experiment.[57,50,51] We review emerging work that has begun to test task-performing computational models with brain and behavioral data. The models include cognitive models, specified at an abstract computational level, whose implementation in biological brains has yet to be explained, and neural network models, which abstract from many features of neurobiology, but could plausibly be implemented with biological neurons. This emerging literature suggests the beginnings of an integrative approach to understanding brain computation, where models are required to perform cognitive tasks, biology provides the admissible component functions, and the computational mechanisms are optimized to explain detailed patterns of brain activity and behavior.

## From experiment toward theory

*Models of connectivity and dynamics*

One path from measured brain activity toward a computational understanding is to model the brain's connectivity and dynamics. Connectivity models go beyond the localization of activated regions and characterize the interactions between regions. Neuronal dynamics can be measured and modeled at multiple scales, from local sets of interacting neurons to whole-brain activity.[58] A first approximation of brain dynamics is provided by the correlation matrix among the measured response time series, which characterizes the pairwise "functional connectivity" between regions. The literature on resting-state networks has explored this approach[59], and linear decompositions of the space-time matrix, such as spatial independent component analysis, similarly capture simultaneous correlations between locations across time.[60]

By thresholding the correlation matrix, the set of regions can be converted into an undirected graph and studied with graph-theoretical methods. Such analyses can reveal "communities" (sets of strongly interacting regions), "hubs" (regions connected to many others) and "rich clubs" (communities of hubs).[61,62] Connectivity graphs can be derived either from anatomical or functional measurements. The anatomical connectivity matrix typically resembles the functional connectivity matrix, because regions interact through anatomical pathways. However, the way anatomical connectivity generates functional connectivity is better modeled by taking local dynamics, delays, indirect interactions, and noise into account.[63] From local neuronal interactions to large-scale spatiotemporal patterns spanning cortex and subcortical regions, generative models of spontaneous dynamics can be evaluated with brain-activity data.

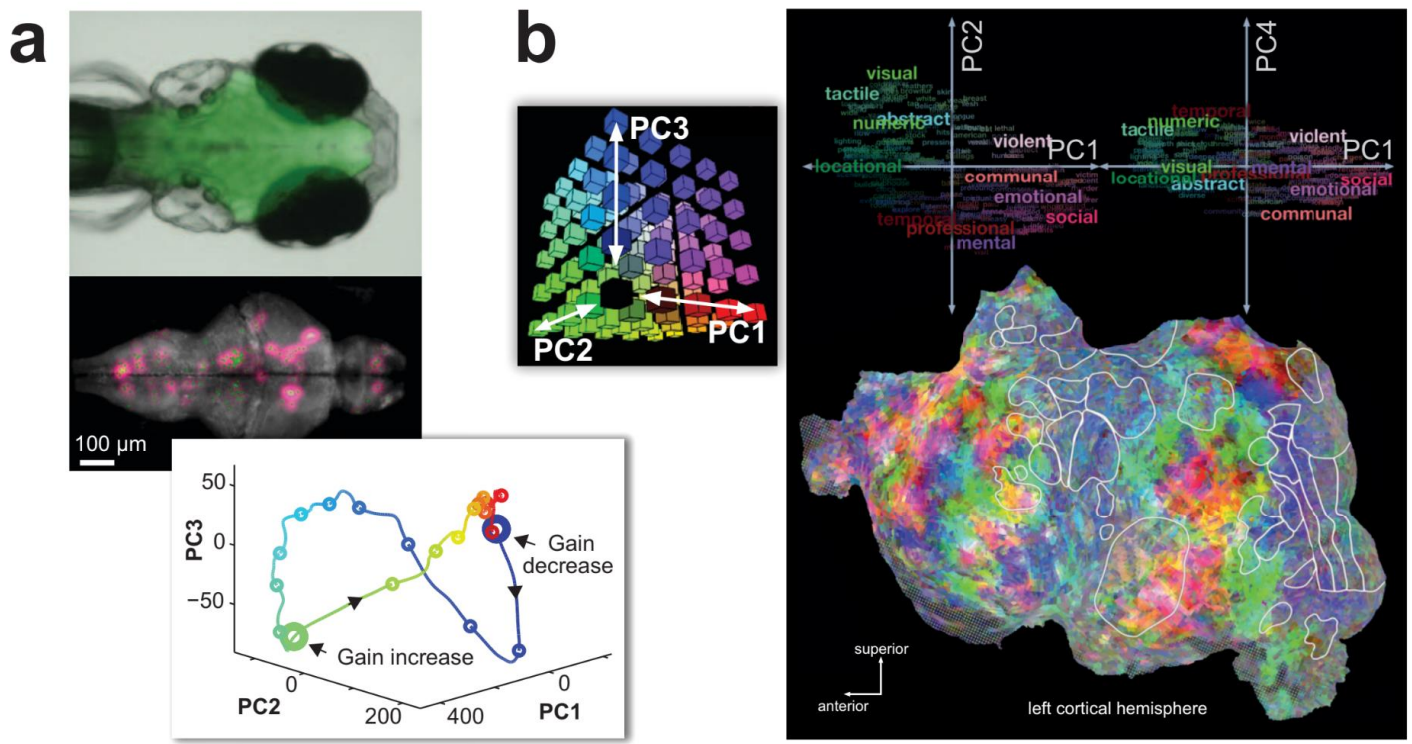

**Figure 1 | Modern imaging techniques provide unprecedentedly detailed information about neuronal activity, but data-driven analyses support only limited insights.** (**a**) Two-photon calcium imaging results from [191] showing single-neuron activity for a large population of cells measured simultaneously in larval zebrafish while the animals interacted with a virtual environment. (**b**) Human fMRI results from [99] reveal a detailed map of semantically selective responses while a subject listened to a story. These results illustrate the power of modern brain-activity measurement techniques and the challenge of drawing insights about brain computation from such data sets. Both studies measured brain activity during complex, time-continuous, naturalistic experience and used linear models (principal component analysis, inset in **a**, bottom) to provide an overall view of the activity patterns and their representational significance.

Effective connectivity analyses take a more hypothesis-driven approach, characterizing the interactions among a small set of regions on the basis of generative models of the dynamics. Dynamic Causal Modeling (DCM), for example, can address questions about causal interactions. The analysis typically focuses on a small set of regions that show task-dependent changes.[64] Granger causality[65] and transfer entropy[66] analyses attempt to infer causal influences based on time series data, by assessing whether considering the past activity of region X boosts prediction of the current activity of region Y, beyond what can be achieved on the basis of the past activity of Y. Effective connectivity analysis usually involves specifying a set of candidate graphical models, each expressing a hypothesis about the causal interactions between regions, and selecting the model that is most consistent with the data.

While activation mapping maps the boxes of cognitive psychology onto brain regions, effective connectivity analyses map the arrows onto pairs of brain regions. Most work in this area has focused on characterizing interactions at the level of the overall activation of a brain region. Like the classical brain mapping approach, these analyses are based on regional-mean activation, measuring correlated fluctuations of overall regional activation rather than the information exchanged between regions. Relating fine-grained pattern information between pairs of regions is also beginning to be explored (e.g. [67]).

Both dynamic models of the whole brain and effective connectivity models of subsets of regions provide potentially meaningful high-level descriptions of the interactions that give rise to brain dynamics. The variation of dynamical states can be studied within individuals (e.g. across states of consciousness[68]) and across individuals (e.g. to understand the changes associated with disorders[69]).

Analyses of effective connectivity and large-scale brain dynamics go beyond generic statistical models like the linear models used in activation and information-based brain mapping in that they are generative models: they can generate data at the level of the measurements and are models *of* brain dynamics. However, they do not capture the information exchanged or the processing occurring in the brain.

*Decoding models*

Another path toward understanding the brain's computational mechanisms is to reveal what information is present in each brain region. Decoding models take brain-activity patterns as input and output what is interpreted as the mental content represented in a brain region.[18,70,71] Decoding can help us go beyond the notion of *activation*, which indicates the involvement of a region in a task, and reveal the *information* present in a region's population activity. When particular content is decodable from a brain region, this indicates the presence of the information. To refer to the brain region as "representing" the content adds a functional interpretation[72]: that the information serves the purpose to inform regions receiving these signals about the content. Ultimately, this interpretation needs to be substantiated by further analyses of how the information affects other regions and behavior.[73,74,75,76]

Decoding has its roots in the neuronal-recording literature[77,40] and has become a popular tool for studying the content of representations in neuroimaging.[78,18,70,79,71,80] In the simplest case, decoding reveals which of two stimuli gave rise to a measured response pattern. The content of the representation can be the identity of a sensory stimulus (to be recognized among a set of alternative stimuli), a stimulus property (such as the orientation of a grating), an abstract variable needed for a cognitive operation, or an action[81]. When the decoder is linear, as is usually the case, the decodable information is in a format that can plausibly be read out by downstream neurons in a single step. Such information is said to be *explicit* in the activity patterns.[73,73,82]

Decoding models enable us to read out the orientation of a visual stimulus from early visual cortex,[16] object categories from ventral stream areas,[15,83] face identity from anterior temporal areas,[28,25] and belief-related decisions from ventromedial prefrontal areas.[84] Decoding studies have also contributed to our understanding of attention,[16,85] and working memory.[86] The literature goes far beyond these examples and has been repeatedly reviewed.[70,80,78,73]

A particularly impressive form of decoding is stimulus reconstruction.[87,88,89] Like ordinary decoding, reconstruction maps brain activity patterns to stimuli. However, the space of stimuli that can be decoded is much more complex (e.g. all natural images) and the decoding model must generalize not just to new measurements for the same stimuli, but to new stimuli. This requires that the model capture something more general about the relationship between stimulus and multivariate response.

Decoding and other types of multivariate pattern analysis have helped reveal the content of regional representations,[78,18,70,71,80] providing evidence that brain-computational models must incorporate. However, the ability to decode particular information does not amount to a full account of the neuronal code: It doesn't specify the representational format (beyond linear decodability) or what other information might additionally be present. Most importantly, decoders do not in general constitute models of brain computation. They reveal aspects of the product, but not the process of brain computation.

*Representational models*

Decoders enable us to test for the presence of particular information in a brain region. Ultimately, we would like to exhaustively characterize a region's representation, explaining its responses to arbitrary stimuli. A full characterization would also define to what extent any variable can be decoded. *Representational models* attempt to make comprehensive predictions about the representational space and therefore provide stronger constraints on the computational mechanism than decoding models.[90,74]

Three types of representational model analysis have been introduced in the literature: encoding models[91,92,93] pattern component models (PCM),[94] and representational similarity analysis (RSA).[95,79,96] These three methods all test hypotheses about the representational space, which are based on multivariate descriptions of the experimental conditions, e.g. a semantic description of a set of stimuli, or the activity patterns across a layer of a neural network model that processes the stimuli.[74]

In encoding models, each voxel's activity profile across stimuli is predicted as a linear combination of the features of the model. In PCM, the distribution of the activity profiles that characterizes the representational space is modeled as a multivariate normal distribution. In RSA, the representational space is characterized by the representational dissimilarities of the activity patterns elicited by the stimuli. Mathematically, these three techniques of representational modeling are closely related in that they all test hypotheses about the representational space defined by the second moment of the activity profiles.[74] All three techniques consider two sets of representational features equivalent if they share the same second moment and, thus, generate the same representational space – providing the same information to downstream regions (but see [97]).

Representational models are often defined on the basis of descriptions of the stimuli, such as labels provided by human observers.[91,98,98,99] In this scenario, a representational model that explains the brain responses in a given region provides, not a brain-computational account, but at least a descriptive account of the representation. Such an account can be a useful stepping stone toward computational theory when the model generalizes to novel stimuli. Importantly, representational models also enable us to adjudicate among brain-computational models, an approach we will return to in the next section.

In this section, we considered three types of model that can help us glean computational insight from brain-activity data. Connectivity models capture aspects of the dynamic interactions between regions. Decoding models enable us to look into brain regions and reveal what might be their representational content. Representational models enable us to test explicit hypotheses that fully characterize a region's representational space. All three types of model can be used to address theoretically motivated questions – taking a hypothesis-driven approach. However, in the absence of task-performing computational models, they are subject to Newell's argument that asking a series of questions might never reveal the computational mechanism underlying the cognitive feat we are trying to explain. These methods fall short of building the bridge all the way to theory, because they do not test mechanistic models that specify precisely how the information processing underlying some cognitive function might work.

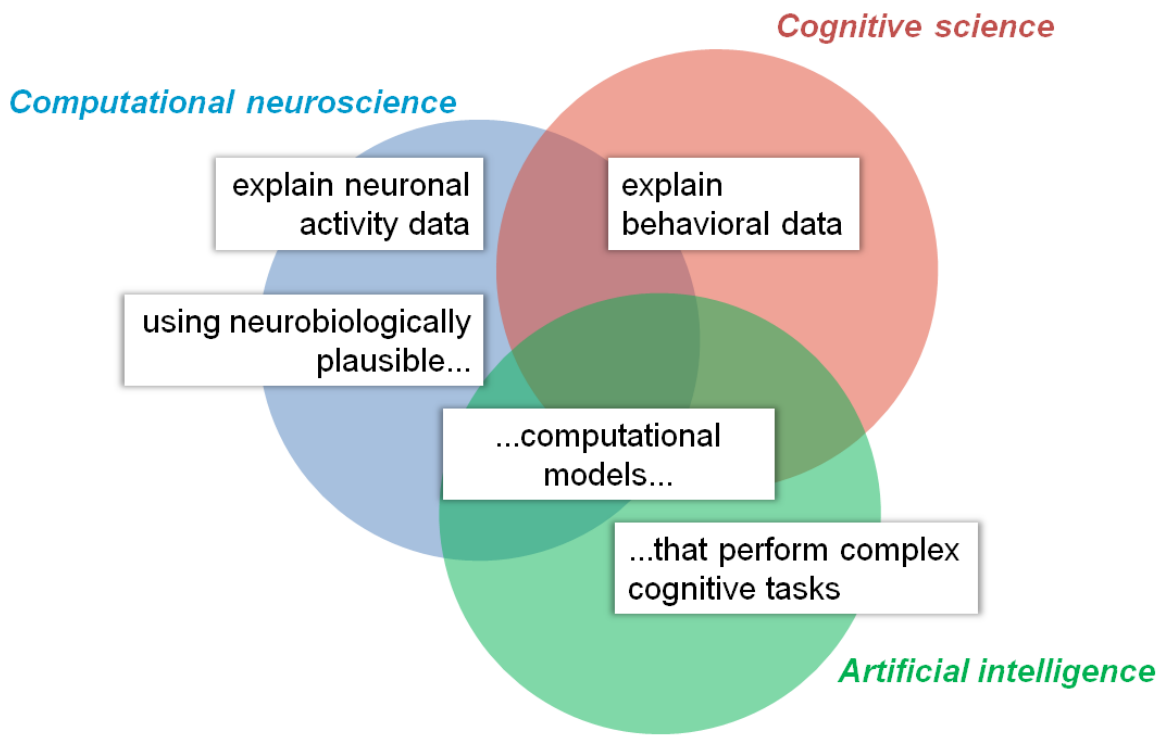

**Figure 2 | What does it mean to understand how the brain works?** The goal of cognitive computational neuroscience is to explain rich measurements of neuronal activity and behavior in animals and humans by means of biologically plausible computational models that perform real-world cognitive tasks. Historically, each of the disciplines (circles) has tackled a subset of these challenges (white labels). Cognitive computational neuroscience strives to meet all the criteria simultaneously.

**Box 1: The many meanings of model**

The word model has many meanings in the brain and behavioral sciences. **Data-analysis models** are generic statistical models that help establish relationships between measured variables. Examples include linear correlation, univariate multiple linear regression for brain mapping, and linear decoding analysis. Effective connectivity and causal-interaction models are, similarly, data-analysis models. They help us infer causal influences and interactions between brain regions. Data-analysis models can serve the purpose to test hypotheses about relationships among variables (e.g. correlation, information, causal influence). They are not models of brain information processing. A **box-and-arrow model**, by contrast, is an information-processing model in the form of labeled boxes that represent cognitive component functions and arrows that represent information flow. In cognitive psychology, such models provided useful, albeit ill-defined, sketches for theories of brain computation. A **word model**, similarly, is a sketch for a theory about brain information processing that is defined vaguely by a verbal description. These are models *of* information processing, but do not *perform* the information processing thought to occur in the brain. An **oracle model** is a model of brain responses (often instantiated in a data-analysis model) that relies on the availability of information, such as semantic properties or category labels, that are not available to the animal whose brain is being modeled. For example, a model of ventral temporal visual responses as a function of an abstract shape description, or as a function of category labels or continuous semantic features constitutes an oracle model, if the model is not capable of computing the shape, category or semantic features from images. An oracle model may provide a useful characterization of the information present in a region and its representational format, without specifying any theory as to how the representation is computed by the brain. A **brain-computational model (BCM)**, by contrast, is a model that mimics the brain information processing underlying the performance of some task at some level of abstraction. In visual neuroscience, for example, an **image-computable model** is a BCM of visual processing that takes image bitmaps as inputs and predicts brain activity and/or behavioral responses. Deep neural nets provide image-computable models of visual processing. However, deep neural nets trained by supervision rely on category-labeled images for training. Because labeled examples are not available (in comparable quantities) during biological development and learning, these models are BCMs of visual processing, but they are not BCMs of development and learning. **Reinforcement learning models** use environmental feedback that is more realistic in quality and can provide BCMs of learning processes. A **sensory encoding model** is a BCM of the computations that transform sensory input to some stage of internal representation. An **internal-transformation model** is a BCM of the transformation of representations between two stages of processing. A **behavioral decoding model** is a BCM of the transformation from some internal representation to a behavioral output. Note that the label BCM indicates merely that the model is intended to capture brain computations at some level of abstraction. A BCM may abstract from biological detail to an arbitrary degree, but must predict some aspect of brain activity and/or behavior. **Psychophysical models** that predict behavioral outputs from sensory input and **cognitive models** that perform cognitive tasks are BCMs formulated at a high level of description. The label BCM does not imply that the model is either plausible or consistent with empirical data. Progress is made by rejecting candidate BCMs on empirical grounds. Like micro-scale **biophysical models**, which capture biological processes that underlie brain computations, and macro-scale **brain-dynamical and causal-interaction models**, BCMs are models of processes occurring in the brain. However, unlike the other types of process model, BCMs *perform the information processing* that is thought to be the function of brain dynamics. Finally, the term model is used to refer to models of the world employed by the brain, as in **model-based reinforcement learning** and **model-based cognition**.

**From theory to experiment**

To build a better bridge between experiment and theory, we first need to fully specify a theory. This can be achieved by defining the theory mathematically and implementing it in a computational model (Box 1). Computational models can reside at different levels of description, trading off cognitive fidelity against biological fidelity (Figure 3). Models designed to capture only neuronal components and dynamics[100] tend to be unsuccessful at explaining cognitive function[101] (Figure 3, horizontal axis). Conversely, models designed to capture only cognitive functions are difficult to relate to the brain (Figure 3, vertical axis). To link mind and brain, models must attempt to capture aspects of both behavior and neuronal dynamics. Recent advances suggest that constraints from the brain can help explain cognitive function,[55,102,103] and vice versa,[50,51] turning the tradeoff into a synergy.

In this section, we focus on recent successes with task-performing models that explain cognitive functions in terms of representations and algorithms. Task-performing models have been central to psychophysics and cognitive science, where they are traditionally tested with behavioral data. An emerging literature is beginning to test task-performing models also with brain-activity data. We will consider two broad classes of model in turn, neural network models and cognitive models.

*Neural network models*

Neural network models (Box 2) have a long history with interwoven strands in multiple disciplines. In computational neuroscience, neural network models, at various levels of biological detail, have been essential to understanding dynamics in biological neural networks and elementary computational functions.[40,41] In cognitive science, they defined a new paradigm for understanding cognitive functions called parallel distributed processing in the 1980s,[6,104] which brought the field closer to neuroscience. In AI, they have recently brought substantial advances in a number of applications,[55,103] ranging from perceptual tasks (such as vision and speech recognition) to symbolic processing challenges (such as language translation), and on to motor tasks (including speech synthesis and robotic control). Neural network models provide a common language for building task-performing models that meet the combined criteria for success of the three disciplines (Figure 2).

Like brains, neural network models can perform feedforward as well as recurrent computations.[50,106] The models driving the recent advances are deep in the sense that they comprise multiple stages of linear-nonlinear signal transformation. Models typically have millions of parameters (the connection weights), which are set so as to optimize task performance. One successful paradigm is supervised learning, where a desired mapping from inputs to outputs learned from a training set of inputs (e.g. images) and associated outputs (e.g. category labels). However, neural network models can also be trained without supervision and can learn complex statistical structure inherent to their experiential data.

The large number of parameters creates unease among researchers who are used to simple models with small numbers of interpretable parameters. However, simple models will never enable us to explain complex feats of intelligence. The history of AI has shown that intelligence requires ample world knowledge, and sufficient parametric complexity to store it. We therefore must engage complex models (Figure 3) and the challenges they pose. One challenge is that the high parameter count renders the models difficult to understand. Because the models are entirely transparent, they can be probed cheaply with millions of input patterns to understand the internal representations, an approach sometimes called "synthetic neurophysiology". To address the concern of overfitting, models are evaluated in terms of their generalization performance. A vision model, for

**Box 2: Neural network models**

The term *neural network model* has come to be associated with a class of model that is inspired by biological neural networks in that each unit combines many inputs and information is processed in parallel through a network. In contrast to biologically detailed models, which may capture action potentials and dynamics in multiple compartments of each neuron, these models abstract from the biological details. However, they can explain certain cognitive functions, such as visual object recognition, and therefore provide an attractive framework for linking cognition to the brain.

A typical unit computes a linear combination of its inputs and passes the result through a static nonlinearity. The output is sometimes interpreted as analogous to the firing rate of a neuron. Even *shallow* networks (those with a single layer of hidden units between inputs and outputs) can approximate arbitrary functions.[105] However, *deep* networks (those with multiple hidden layers) can more efficiently capture many of the complex functions needed in real-world tasks. Many applications, e.g. in computer vision, use feedforward architectures. However, recurrent neural networks, which reprocess the outputs of their units and generate complex dynamics, have brought additional engineering advances[106] and better capture the recurrent signaling in brains.[107,108,48,109] Whereas feedforward networks are universal function approximators, recurrent networks are universal approximators of dynamical systems.[110] Recurrent processing enables a network to recycle its limited computational resources through time, so as to perform more complex sequences of computations. Recurrent networks can represent the recent stimulus history in a dynamically compressed format, providing the temporal context information needed for current processing. As a result, recurrent networks can recognize, predict, and generate dynamical patterns.

Neural network models learn their parameters. Both feedforward and recurrent networks are defined by their architecture and the setting of the connection weights. One way to set the weights is through iterative small adjustments that bring the output closer to some desired output (supervised learning). Each weight is adjusted in proportion to the reduction in the error that a small change to it would yield. This method is called *gradient descent*, because it produces steps in the space of weights along which the error declines most steeply. Gradient descent can be implemented using *backpropagation*, an efficient algorithm for computing the derivative of the error function with respect to each weight.

Whether the brain uses an algorithm like backpropagation for learning is controversial. Several biologically plausible implementations of backpropagation or closely related forms of supervised learning have been suggested.[111,112,113] Supervision signals might be generated internally[114] on the basis of the context provided by multiple sensory modalities, on the basis of the dynamic refinement of representations over time, as more evidence becomes available from the senses and from memory[115], and on the basis of internal and external reinforcement signals arising in interaction with the environment.[116] Reinforcement learning[54] and unsupervised learning of neural network parameters[117,118] are areas of rapid current progress.

Neural network models have demonstrated that taking inspiration from biology can yield breakthroughs in AI. It seems likely that the quest for models that can match human cognitive abilities will draw us deeper into the biology.[185] The abstract neural network models currently most successful in engineering could be implemented with biological hardware. However, they only use a small subset of the dynamical components of brains. Neuroscience has described a rich repertoire of dynamical components, including action potentials,[165] canonical microcircuits[186], dendritic dynamics,[187,111,113] and network phenomena,[40] such as oscillations,[188] which may have computational functions. Integrating these dynamical components into computational models designed to perform meaningful tasks promises to reveal their computational function in the brain and may drive further advances in AI.

example, will be evaluated in terms of its ability to predict neural activity and behavioral responses for images it has not been trained on.

Several recent studies have begun to test neural network models as models of brain information processing.[50,51] These studies used deep convolutional neural network models trained to recognize objects in images to predict brain representations of images in the primate ventral visual stream. Results have shown that the internal representations of deep convolutional neural networks provide the best current models of representations of visual images in inferior temporal (IT) cortex in humans and monkeys.[119,120,121] When comparing large numbers of models, those that were optimized to perform the task of object classification better explained the IT representation.[119,120]

Early layers of deep neural networks trained to recognize objects contain representations resembling early visual cortex.[120,122] As we move gradually along the ventral visual stream, higher layers of the neural networks gradually come to provide a better basis for explaining the representations.[122,123,124] Higher layers of deep convolutional neural networks also resemble the IT cortical representation in that both enable the decoding of object position, size, and pose, along with the category of the object.[125] In addition to testing these models by predicting brain-activity data, the field has begun to test them by predicting behavioral responses reflecting perceived shape[126] and object similarity.[127]

*Cognitive models*

Models at the cognitive level enable researchers to envision the information processing without simultaneously having to tackle its implementation with neurobiologically plausible components. This enables progress on domains of higher cognition, where neural network models still fall short. Moreover, a cognitive model may provide a useful abstraction, even when a process can also be captured with a neural network model.

Neuroscientific explanations now dominate for functional components closer to the periphery of the brain where sensory and motor processes connect the animal to its environment. However, much of higher-level cognition has remained beyond the reach of neuroscientific accounts and neural network models. From its early days, cognitive science has built task-performing models using symbolic representations. These models explained high-level cognitive processes including judgments, problem solving, and planning. To illustrate some of the unique contributions of cognitive models, we briefly discuss three classes of cognitive model: production systems, reinforcement learning models, and Bayesian cognitive models.

Production systems provide an early example of a class of cognitive models that can explain reasoning and problem solving. These models use rules and logic, and are symbolic in that they operate on symbols rather than sensory data and motor signals. They capture cognition, rather than perception and motor control, which ground cognition in the physical environment. A *production* is a cognitive action triggered according to an if-then rule. A set of such rules specifies the conditions ("if") under which each of a range of productions ("then") is to be executed. The conditions refer to current goals and knowledge in memory. The actions can modify the internal state of goals and knowledge. For example, a production may create a subgoal or store an inference. If conditions are met for multiple rules, a conflict resolution mechanism chooses one production. A model specified using this formalism will generate a sequence of productions, which may to some extent resemble our conscious stream of thought while working toward some cognitive goal. The formalism of production systems also provides a universal computational architecture.[133] Production systems such as ACT-R [5] were originally developed under the guidance of behavioral data. More recently such models have also begun to be tested in terms of their ability to predict regional-mean fMRI activation time courses.[134]

Reinforcement learning (RL) models capture how an agent can learn to maximize its long-term cumulative reward through interaction with its environment.[128] As in production systems, RL models often assume that the agent has perception and motor modules that enable the use of discrete symbolic representations of states and actions. The agent chooses actions, observes resulting states of the environment, receives rewards along the way, and learns to improve its behavior. The agent may learn a *value function* associating each state with its expected cumulative reward. If the agent can predict which state each action leads to, and if it knows the values of those states, then it can choose the most promising action. The agent may also learn a *policy* that associates each state directly with promising actions. The choice of action must balance exploitation (which brings short-term reward) and exploration (which benefits learning and brings long-term reward).

The field of RL explores algorithms that define how to act and learn, so as to maximize cumulative reward. With roots in psychology and neuroscience, RL theory is now an important field of machine learning and AI. It provides a very general perspective on control that includes the classical techniques dynamic programming, Monte Carlo and exhaustive search as limiting cases, and can handle challenging scenarios, where the environment is stochastic and only partially observed, and its causal mechanisms are unknown.

An agent might exhaustively explore an environment and learn the most promising action to take in any state by trial and error (model-free control). This would require sufficient time to learn, enough memory, and an environment that does not kill the agent prematurely. Biological organisms, however, have limited time to learn and limited memory, and they are not free to explore interactions that might kill them. Under these conditions, an agent might do better to build a model of its environment. A model can compress and generalize experience to enable intelligent action in novel situations (model-based control). Model-free methods are computationally efficient (mapping from states to values or directly to actions), but statistically inefficient (learning takes long); model-based methods are more statistically efficient, but may require prohibitive amounts of computation (to simulate possible futures).[135]

Until experience is sufficient to build a reliable model, an agent might do best to simply store episodes and revert to paths of action that have met with success in the past (episodic control).[136,137] Storing episodes preserves sequential dependency information important for model building. Moreover, episodic control enables the agent to exploit such dependencies even before understanding the causal mechanism supporting a successful path of action.

The brain is capable of each of these three modes of control (model-free, model-based, episodic)[138] and appears to combine their advantages using an algorithm that has yet to be discovered. AI and computational neuroscience share the goal to discover this algorithm,[139,140,141,54,53,130,142,135] although they approach this goal from different angles. This is an example of how a cognitive challenge can motivate the development of formal models and drive progress in AI and neuroscience.

A third and critically important class of cognitive model is that of Bayesian models (Box 3).[33,35,36,34,143,144,145,32] Bayesian inference provides an essential normative perspective on cognition. It tells us what a brain *should* compute for an animal to behave optimally. Perceptual inference, for example, should consider the current sensory data in the context of prior beliefs. Bayesian inference simply refers to combining the data with prior beliefs according to the rules of probability.

Bayesian models have contributed to our understanding of basic sensory and motor processes.[33,35,36,34] They have also provided insights into higher cognitive processes of judgment and decision making, explaining classical cognitive biases[154] as the product of prior assumptions, which may be incorrect in the experimental task, but correct and helpful in the real world.

With Bayesian nonparametric models, cognitive science has begun to explain more complex cognitive abilities. Consider the human ability to induce a new object category from a single example. Such inductive inference requires prior knowledge of a kind not captured by current feedforward neural network models.[156] To induce a category, we rely on an understanding of the object, of the interactions among its parts, of how they give rise to its function. In the Bayesian cognitive perspective, the human mind, from infancy, builds mental models of the world.[2] These models may not only be generative models in the probabilistic sense, but may be causal and compositional, supporting mental simulations of processes in the world using elements that can be re-composed to generalize to novel and hypothetical scenarios.[145,2,157] This modeling approach has been applied to our reasoning about the physical[158,157,159] and even the social[160] world.

Generative models are an essential ingredient of general intelligence. An agent attempting to model all relationships among the data does not require external supervision or reinforcement to learn. Unsupervised learning greatly expands the amount of exploitable data, enabling the agent to mine all its experiences to understand its environment and itself. In particular, causal models of processes in the world (how objects cause images, how the present causes the future) can give an agent a deeper understanding and, thus, a better basis for inferences and actions.

The representation of probability distributions in neuronal populations has been explored theoretically and experimentally.[161,162] However, relating Bayesian inference and learning, especially structure learning in nonparametric models, to its implementation in the brain remains challenging.[163] As theories of brain computation, approximate inference algorithms like sampling may explain cortical feedback signals and activity correlations.[144,164,165,166,167] Moreover, the corners cut by the brain for computational efficiency, the approximations, may explain human deviations from statistical optimality. In particular, cognitive experiments have revealed signatures of sampling[168] and amortized inference[169] in human behavior.

Cognitive models, including the three classes highlighted here, decompose cognition into meaningful functional components. By declaring their models independent of the implementation in the brain, cognitive scientists are able to address high-level cognitive processes[144,32,145] that are beyond the reach of current neural networks (but see [170]). Cognitive models are essential for cognitive computational neuroscience because they enable us to see the whole as we attempt to understand the roles of the parts.

The fact that cognitive models are not restricted to neurobiologically plausible components is both a strength and a weakness. It enables them to capture complex cognitive phenomena, but makes them difficult to relate to neural network models and measurements of brain activity. Cognitive models are sometimes tested with human functional imaging data. Such tests compare predictions about the time course of certain signals of interest at the coarse scale of regional-mean activation.[134,171,138] A challenge for the future is to test cognitive models at the level of the representational content being processed – as is beginning to happen for neural networks in the framework of representational models.[50,51] For example, we should be able to predict representations of particular pieces of cognitive content, such as the current or currently simulated state of the environment. To the extent that we can make neural predictions from cognitive algorithms or, conversely, derive algorithmic models from neural dynamics, we will be able to leverage cognitive models to understand the brain.

Bayesian cognitive models have recently flourished in interaction with machine learning and statistics. Early work used generative models with a fixed structure that were flexible only with respect to a limited set of parameters. Modern generative models can grow in complexity with the data and discover their inherent structure.[145] They are called *nonparametric*, because they are not limited by a predefined finite set of parameters.[155] Their parameters can grow in number without any predefined bound.

## Box 3: Bayesian cognitive models

*Bayesian cognitive models* are motivated by the assumption that the brain approximates the statistically optimal solution to a task. The statistically optimal way to make inferences and decide what to do is to interpret the current evidence in light of all available prior knowledge using the rules of probability. Consider the case of visual perception. The retinal signals reflect the objects in the world, which we would like to recognize. To infer the objects, we should consider what configurations of objects we deem possible and how well each explains the image. Our prior beliefs are represented by a *generative model* that captures the probability of each configuration of objects and the probabilities with which a given configuration would produce different retinal images.

More formally, a Bayesian model of vision might use a generative model of the joint distribution p(**d**, **c**) of the sensory data **d** (the image) and the causes in the world **c** (the configuration of surfaces, objects, and light sources to be inferred).[146] The joint distribution p(**d**, **c**) equals the product of the *prior* p(**c**) over all possible configurations of causes and the *likelihood* p(**d**|**c**), the probability of a particular image given a particular configuration of causes. A *prescribed* model for p(**d**|**c**) would enable us to evaluate the likelihood, the probability of a specific image **d** given causes **c**. Alternatively, we might have an *implicit* model for p(**d**|**c**) in the form of a stochastic mapping from causes **c** to data **d** (images). Such a model would generate natural images. Whether prescribed or implicit, the model of p(**d**|**c**) captures how the causes in the world create the image, or at least how they relate to the image. Visual recognition amounts to computing the posterior p(**c**|**d**), the probability distribution over the causes given a particular image. The posterior p(**c**|**d**) reveals the causes **c** as they would have to exist in the world to explain the sensory data **d**.[147] A model computing p(**c**|**d**) is called a *discriminative model*, because it discriminates among images – here mapping from effects (the image) to the causes. The inversion mathematically requires a prior p(**c**) over the latent causes. The prior p(**c**) can constrain the interpretation and help reduce the ambiguity resulting from the multiple configurations of causes that can account for any image.

Basing the inference of the causes **c** on a generative model of p(**d**, **c**) that captures all available knowledge and uncertainty is statistically optimal, but computationally challenging (i.e. it may require more neurons or time than the animal can use).. Ideally, the generative model p(**d**, **c**) implicit to the inference p(**c**|**d**) should capture our knowledge not just about image formation, but also the things in the world and their interactions, and our uncertainties about these processes. One challenge is to learn a generative model from sensory data. We need to represent the learned knowledge and the remaining uncertainties. If the generative model is misspecified, then the inference will not be optimal. For real-world tasks, some degree of misspecification of the model is inevitable. For example, the generative model may contain an overly simplified version of the image-generation process. Another challenge is the computation of the posterior p(**c**|**d**). For realistically complex generative models, the inference may require computation-intensive iterative algorithms such as *Markov Chain Monte Carlo* (MCMC), *belief propagation* (BP), or *variational inference*. The brain's compromise between statistical and computational efficiency,[149,150,151] may involve learning fast feedforward recognition models that speed up frequent component inferences, crystallizing conclusions that are costly to fluidly derive with iterative algorithms. This is known as amortized inference.[152,153]

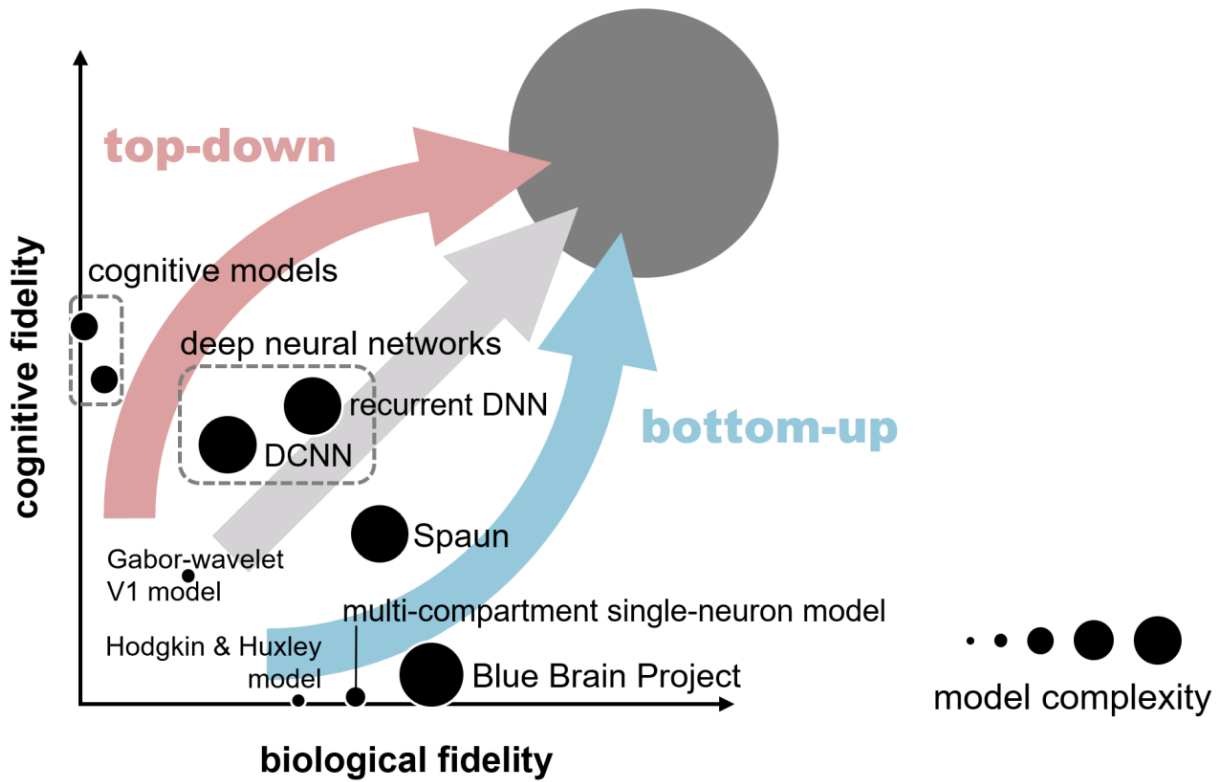

**Figure 3 | The space of process models.** Models of the processes taking place in the brain can be defined at different levels of description and can vary in their parametric complexity (dot size) and in their biological (horizontal axis) and cognitive (vertical axis) fidelity. Theoreticians approach modeling with a range of primary goals. The bottom-up approach to modeling (blue arrow) aims first to capture characteristics of biological neural networks, such as action potentials and interactions among multiple compartments of single neurons. This approach disregards cognitive function, so as to focus on understanding the emergent dynamics of small parts of the brain, such as cortical columns and areas, and to reproduce biological network phenomena, such as oscillations. The top-down approach (red arrow) aims first to capture cognitive functions at the algorithmic level. This approach disregards the biological implementation, so as to focus on decomposing the information processing underlying task performance into its algorithmic components. The two approaches form the extremes of a continuum of paths toward the common goal to explain how our brains give rise to our minds. Overall there is tradeoff (negative correlation) between cognitive and biological fidelity. However, the tradeoff can turn into a synergy (positive correlation) when cognitive constraints illuminate biological function, and when biology inspires models that explain cognitive feats. Because intelligence requires rich world knowledge, models of human brain information processing will have high parametric complexity (large dot in the upper right corner). Even if models that abstract from biological details can explain task performance, biologically detailed models will still be needed to explain the neurobiological implementation. This diagram is a conceptual cartoon that can help us understand the relationships between models and appreciate their complementary contributions. However, it is not based on quantitative measures of cognitive fidelity, biological fidelity, and model complexity. Definitive ways to measure each of the three variables have yet to be developed. Figure inspired by [192].

> **Box 4: Why do cognitive science, computational neuroscience, and AI need each other?**
>
> **Cognitive science** needs computational neuroscience, not merely to explain the implementation of cognitive models in the brain, but also to discover the algorithms. For example, the dominant models of sensory processing and object recognition are brain-inspired neural networks, whose computations are not easily captured at a cognitive level. Recent successes with Bayesian nonparametric models do not yet in general scale to real-world cognition. Explaining the computational efficiency of human cognition and predicting detailed cognitive dynamics and behavior could benefit from studying brain-activity dynamics. Explaining behavior is essential, but behavioral data alone provide insufficient constraints for complex models. Brain data can provide rich constraints for cognitive algorithms if leveraged appropriately. Cognitive science has always progressed in close interaction with artificial intelligence. The disciplines share the goal of building task-performing models and thus rely on common mathematical theory and technologies.
>
> **Computational neuroscience** needs cognitive science to challenge it to engage higher-level cognition. At the experimental level, the tasks of cognitive science enable computational neuroscience to bring cognition into the lab. At the level of theory, cognitive science challenges computational neuroscience to explain how the neurobiological dynamical components it studies contribute to cognition and behavior. Computational neuroscience needs AI, and in particular machine learning, to provide the theoretical and technological basis for modeling cognitive functions with biologically plausible dynamical components.
>
> **Artificial intelligence** needs cognitive science to guide the engineering of intelligence. Cognitive science's tasks can serve as benchmarks for AI systems, building up from elementary cognitive abilities to artificial general intelligence. The literatures on human development and learning provide an essential guide to what is possible for a learner to achieve and what kinds of interaction with the world can support the acquisition of intelligence. AI needs computational neuroscience for algorithmic inspiration. Neural network models are an example of a brain-inspired technology that is unrivalled in several domains of AI. Taking further inspiration from the neurobiological dynamical components (e.g. spiking neurons, dendritic dynamics, the canonical cortical microcircuit, oscillations, neuromodulatory processes) and the global functional layout of the human brain (e.g. subsystems specialized for distinct functions, including sensory modalities, memory, planning, motor control) might lead to additional AI breakthroughs. Machine learning draws from separate traditions in statistics and computer science, which have optimized statistical and computational efficiency, respectively. The integration of computational and statistical efficiency is an essential challenge in the age of big data. The brain appears to combine computational and statistical efficiency and understanding its algorithm might boost machine learning.

## Looking ahead

*Bottom up and top down*

The brain seamlessly merges bottom-up discriminative and top-down generative computations in perceptual inference, and model-free and model-based control. Brain science likewise needs to integrate its levels of description and progress both bottom-up and top-down, so as to explain task performance on the basis of neuronal dynamics and provide a mechanistic account of how the brain gives rise to the mind.

Bottom-up visions, proceeding from detailed measurements toward an understanding of brain computation, have been prominent and have driven the most important recent funding initiatives. The European Human

Brain Project and the U.S. BRAIN Initiative are both motivated by bottom-up visions, in which an understanding of brain computation is achieved by measuring and modeling brain dynamics with a focus on the circuit level. The BRAIN Initiative seeks to advance technologies for measuring and manipulating neuronal activity. The Human Brain Project attempts to synthesize neuroscience data in biologically detailed dynamic models. Both initiatives proceed primarily from experiment toward theory, and from the cellular level of description to larger-scale phenomena.

Measuring large numbers of neurons simultaneously and modeling their interactions at the circuit level will be essential. The bottom-up vision is grounded in the history of science. Microscopes and telescopes, for example, have brought scientific breakthroughs. However, it is always in the context of prior theory (generative models of the observed processes) that better observations advance our understanding. In astronomy, for example, the theory of Copernicus guided Galileo in interpreting his telescopic observations.

Understanding the brain requires that we develop theory and experiment in tandem, and complement the bottom-up, data-driven approach by a top-down, theory-driven approach that starts with behavioral functions to be explained.[172,173] Unprecedentedly rich measurements and manipulations of brain activity will drive theoretical insight when they are used to adjudicate between brain-computational models that pass the first test of being able to perform a function that contributes to the behavioral fitness of the organism. The top-down approach, therefore, is an essential complement to the bottom-up approach toward understanding the brain (Figure 3).

*Integrating Marr's levels*

Marr (1982) offered a distinction of three levels of analysis: (1) computational theory, (2) representation and algorithm, and (3) neurobiological implementation.[174] Cognitive science starts from computational theory, decomposing cognition into components and developing algorithms from the top down. Computational neuroscience proceeds from the bottom up, composing neuronal building blocks into representations and algorithms thought to be useful components in the context of the brain's overall goal to control behavior. AI builds representations and algorithms that combine simple components to implement complex feats of intelligence. All three disciplines, thus, converge on the algorithms and representations of the brain and mind, contributing complementary constraints.

Marr's levels provide a useful guide to the challenge of understanding the brain. However, they should not be taken to suggest that cognitive science need not consider the brain, or that computational neuroscience need not consider cognition (Box 4). Marr was inspired by computers, which are designed by human engineers to precisely conform to high-level algorithmic descriptions. This enables the engineers to abstract from the circuits when designing the algorithms. Even in computer science, however, certain aspects of the algorithms depend on the hardware, such as its parallel processing capabilities. Brains differ from computers in ways that exacerbate this dependence. Brains are the product of evolution and development, processes that are not constrained to generate systems whose behavior can be perfectly captured at some abstract level of description. It may therefore not be possible to understand cognition without considering its implementation in the brain or, conversely, to make sense of neuronal circuits, except in the context of the cognitive functions they support.

For an example of a challenge that transcends the disciplines, consider a child seeing an escalator for the first time. She will rapidly recognize people on steps traveling upward obliquely. She might think of it as a moving staircase and imagine riding on it, being lifted one story without exerting any effort. She might infer its function and form a new concept on the basis of a single experience, before ever learning the word "escalator".

Deep neural network models provide a biologically plausible account of the rapid recognition of the elements of the visual experience (people, steps, oblique upward motion, handrail). They can explain the computationally efficient pattern recognition component.[55] However, they cannot explain yet how the child understands the relationships among the elements, the physical interactions of the objects, the people's goal to go up, and the function of the escalator, or how she can imagine the experience and instantly form a new concept.

Bayesian nonparametric models explain how deep inferences and concept formation from single experiences are even possible. They may explain the brain's stunning statistical efficiency, its ability to infer so much from little data by building generative models that provide abstract prior knowledge.[145] However, current inference algorithms require large amounts of computation and, as a result, do not yet scale to real-world challenges like forming the new concept "escalator" from a single visual experience.

On a 20-Watt power budget, the brain's algorithms combine statistical and computational efficiency in ways that are beyond current AI of either the Bayesian or the neural network variety. However, recent work in AI and machine learning has begun to explore the intersection between Bayesian inference and neural network models, combining the statistical strengths of the former (uncertainty representation, probabilistic inference, statistical efficiency) with the computational strengths of the latter (representational learning, universal function approximation, computational efficiency).[175,170,117] In cognitive science, analysis-by-synthesis models of visual recognition, which use iterative fitting of a generative graphics model to the image, have been complemented with discriminative neural networks that provide rapid feedforward estimates.[27,176,177] Generative neural network models are also a long-standing[178, 179,180] and now quickly growing area of research.[118,117] The intersection between probabilistic inference and neural network models seems poised for further breakthroughs that could impact brain and cognitive theory as well as AI.

More generally, neural network models can be integrated with other techniques that have complementary strengths. For example, neural network models can be trained by RL[54] and integrated with classical AI techniques.[181,130] Neural networks can also be enhanced with more stable memory components.[182,183] This addresses the need for memory at different time scales and is inspired by the theory of complementary learning systems in the brain.[184]

Integrating all three of Marr's levels will require close collaboration among researchers with a wide variety of expertise. It is difficult for any single lab to excel at neuroscience, cognitive science, and AI-scale computational modeling. We therefore need collaborations between labs with complementary expertise. In addition to conventional collaborations, an open science culture, in which components are shared between disciplines, can help us integrate Marr's levels. Shareable components include cognitive tasks, brain and behavioral data, computational models, and tests that evaluate models by comparing them to biological systems (Box 5).

The study of the mind and brain is entering a particularly exciting phase. Recent advances in computer hardware and software enable AI-scale modeling of the mind and brain. If cognitive science, computational neuroscience, and artificial intelligence can come together, we might be able to explain human cognition with neurobiologically plausible computational models.

## Box 5: Shareable tasks, data, models, and tests – a new culture of multidisciplinary collaboration

Neurobiologically plausible models that explain cognition will have substantial parametric complexity. Building and evaluating such models will require machine learning, and big brain and behavioral data sets. Traditionally, each lab has developed its own tasks, data sets, models, and tests with a focus on the goals of its own discipline. To scale these efforts up to the challenge, we will need to develop tasks, data, models, and tests that are relevant across the three disciplines and shared among labs (Figure). A new culture of collaboration will assemble big data and big models by combining components from different labs. To meet the conjoined criteria for success of cognitive science, computational neuroscience, and artificial intelligence, the best division of labor might cut across the traditional disciplines.

**Tasks**: By designing experimental tasks, we carve up cognition into components that can be quantitatively investigated. A task is a controlled environment for behavior. It defines the dynamics of a task "world" that provides sensory input (e.g. visual stimuli) and captures motor output (e.g. button press, joystick control, or higher-dimensional limb or whole-body control). Tasks drive the acquisition of brain and behavioral data and the development of AI models, providing well-defined challenges and quantitative performance benchmarks for comparing models. The ImageNet tasks,[189] for example, have driven substantial progress in computer vision. Tasks should be designed and implemented such that they can readily be used in all three disciplines to drive data acquisition and model development (related developments include: OpenAI's Gym, https://gym.openai.com, and Universe, https://universe.openai.com; and DeepMind's Lab[193]). The spectrum of useful tasks includes classical psychophysical tasks employing simple stimuli and responses as well as interactions in virtual realities. As we engage all aspects of the human mind, our tasks will need to simulate natural environments and will come to resemble computer games. This may bring the added benefit of mass participation and big behavioral data, especially when tasks are performed via the web.[190]

**Data**: Behavioral data acquired during task performance provides overall performance estimates and detailed signatures of success and failure, of reaction times and movement trajectories. Brain-activity measurements characterize the dynamic computations underlying task performance. Anatomical data can characterize the structure and connectivity of the brain at multiple scales. Structural brain data, functional brain data, and behavioral data will all be essential for constraining computational models.

**Models**: Task-performing computational models can take sensory inputs and produce motor outputs, so as to perform experimental tasks. AI-scale neurobiologically plausible models can be shared openly and tested in terms of their task performance and in terms of their ability to explain a variety of brain and behavioral data sets, including new data sets acquired after definition of the model. Initially, many models will be specific to small subsets of tasks. Ultimately, models must generalize across tasks.

**Tests**: To assess the extent to which a model can explain brain information processing during a particular task, we need tests that compare models and brains on the basis of brain and behavioral data. Every brain is idiosyncratic in its structure and function. Moreover, for a given brain, every act of perception, cognition, and action is unique in time and cannot be repeated precisely because it permanently changes the brain in question. These complications make it challenging to compare brains and models. We must define the summary statistics of interest and the correspondence mapping between model and brain in space and time at some level of abstraction. Developing appropriate tests for adjudicating among models and determining how close we are to understanding the brain is not merely a technical challenge of statistical inference. It is a conceptual challenge fundamental to theoretical neuroscience.

The interaction among labs and disciplines can benefit from adversarial cooperation.[118] Cognitive researchers who feel that current computational models fall short of explaining an important aspect of cognition are challenged to design shareable tasks and tests that quantify these shortcomings, and to provide human behavioral data to set the bar for AI models. Neuroscientists who feel that current models do not explain brain information processing are challenged to share brain-activity data acquired during task performance and tests comparing activity patterns between brains and models to quantify the shortcomings of the models. Although we will have a plurality of definitions of success, translating these into quantitative measures of the quality of a model is essential and could drive progress in cognitive computational neuroscience as well as engineering.

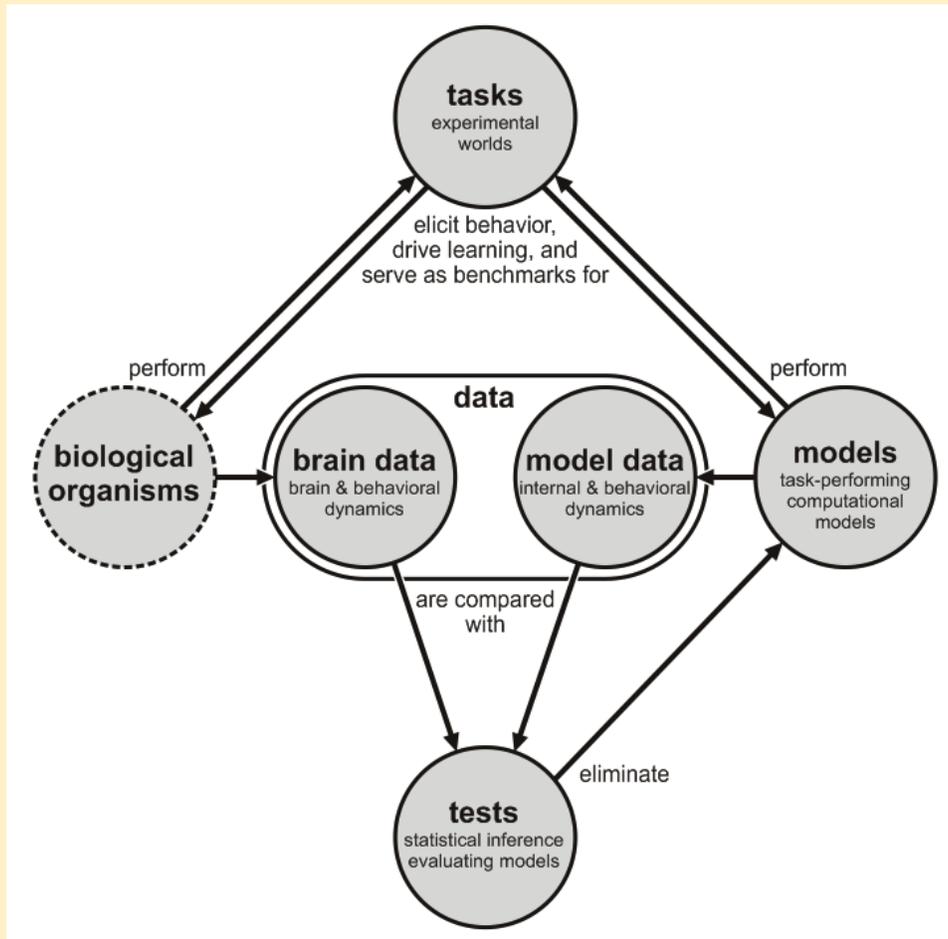

**Figure | Interactions among shareable components.** Tasks, data, models, and tests are components (gray nodes) that lend themselves to sharing among labs and across disciplines, to enable collaborative construction and testing of big models driven by big brain and behavioral data sets assembled across labs.

**Acknowledgements**

This paper benefited from discussions in the context of the new conference *Cognitive Computational Neuroscience*, which had its inaugural meeting in New York City in September 2017.[194] We are grateful in particular to Thomas Naselaris, Kendrick Kay, Konrad Kording, Daphna Shohamy, Russ Poldrack, Jörn Diedrichsen, Matthias Bethge, Robert Mok, Tim Kietzmann, Katherine Storrs, Marieke Mur, Tal Golan, Mate Lengyel, Michael Shadlen, Daniel Wolpert, Aude Oliva, Daniel Yamins, Jonathan Cohen, James DiCarlo, Talia Konkle, Josh McDermott, Nancy Kanwisher, Sam Gershman, and Josh Tenenbaum for inspiring discussions.

# References

1. Newell, A. You can't play 20 questions with nature and win: Projective comments on the papers of this symposium. *Technical Report, School of Computer Science. Carnegie Mellon University* (1973).

2. Lake, B. M., Ullman, T. D., Tenenbaum, J. B. & Gershman, S. J. Building machines that learn and think like people. *Behavioral and Brain Sciences* **40,** (2017).

3. Kriegeskorte, N. & Mok, R. M. Building machines that adapt and compute like brains. *Behavioral and Brain Sciences* **40,** (2017).

4. Simon, H. A. & Newell, A. Human problem solving: The state of the theory in 1970. *American Psychologist* **26,** 145–159 (1971).

5. Anderson, J. R. *The architecture of cognition.* (Harvard University Press, 1983).

6. McClelland, J. L. & Rumelhart, D. E. Parallel distributed processing. *Cambridge,MA, MIT Press* **1,2,** (1987).

7. Berger, H. Über das Elektrenkephalogramm des Menschen. *Archiv fur Psychiatrie und Nervenkrankheiten* **87,** 527–570 (1929).

8. Cohen, D. Magnetoencephalography: evidence of magnetic fields produced by alpha rhythm currents. *Science* **161,** 784–786 (1968).

9. Ogawa, S. Intrinsic signal changes accompanying sensory stimulation: functional brain mapping with magnetic resonance imaging. *Proceedings of the National Academy of Sciences* **89,** 5951–5955 (1992).

10. Gazzaniga et al. *The cognitive neurosciences.* (MIT Press, 2004).

11. Fodor, J. A. Précis of The Modularity of Mind. *Behavioral and Brain Sciences* **8,** 1 (1985).

12. Friston, K. J. *et al.* Statistical parametric maps in functional imaging: A general linear approach. *Human Brain Mapping* **2,** 189–210 (1994).

13. Chklovskii, D. B. & Koulakov, A. A. MAPS IN THE BRAIN: What Can We Learn from Them? *Annual Review of Neuroscience* **27,** 369–392 (2004).

14. Poldrack, R. A. & Yarkoni, T. From Brain Maps to Cognitive Ontologies: Informatics and the Search for Mental Structure. *Annual Review of Psychology* **67,** 587–612 (2016).

15. Haxby, J. V. Distributed and Overlapping Representations of Faces and Objects in Ventral Temporal Cortex. *Science* **293,** 2425–2430 (2001).

16. Kamitani, Y. & Tong, F. Decoding the visual and subjective contents of the human brain. *Nature Neuroscience* **8,** 679–685 (2005).

17. Kriegeskorte, N., Goebel, R. & Bandettini, P. Information-based functional brain mapping. *Proc. Natl. Acad. Sci. U.S.A* **103,** 3863–3868 (2006).

18. Haynes, J.-D. & Rees, G. Decoding mental states from brain activity in humans. *Nature Reviews Neuroscience* **7,** 523–534 (2006).

19. Szucs, D. & Ioannidis, J. P. A. Empirical assessment of published effect sizes and power in the recent cognitive neuroscience and psychology literature. *PLOS Biology* **15,** e2000797 (2017).

20. Kriegeskorte, N., Simmons, W. K., Bellgowan, P. S. F. & Baker, C. I. Circular analysis in systems neuroscience: the dangers of double dipping. *Nature Neuroscience* **12,** 535–540 (2009).

21. Kanwisher, N., McDermott, J. & Chun, M. M. The fusiform face area: a module in human extrastriate cortex specialized for face perception. *J. Neurosci.* **17,** 4302–4311 (1997).

22. Grill-Spector, K., Knouf, N. & Kanwisher, N. The fusiform face area subserves face perception, not generic within-category identification. *Nature Neuroscience* **7,** 555–562 (2004).

23. Tsao, D. Y., Freiwald, W. A., Knutsen, T. A., Mandeville, J. B. & Tootell, R. B. H. Faces and objects in macaque cerebral cortex. *Nature Neuroscience* **6,** 989–995 (2003).

24. Tsao, D. Y. A Cortical Region Consisting Entirely of Face-Selective Cells. *Science* **311,** 670–674 (2006).

25. Freiwald, W. A. & Tsao, D. Y. Functional Compartmentalization and Viewpoint Generalization Within the Macaque Face-Processing System. *Science* **330,** 845–851 (2010).

26. Grill-Spector, K., Weiner, K. S., Kay, K. & Gomez, J. The Functional Neuroanatomy of Human Face Perception. *Annual Review of Vision Science* **3,** 167–196 (2017).

27. Yildirim, I. *et al.* Efficient and robust analysis-by-synthesis in vision: A computational framework, behavioral tests, and modeling neuronal representations. *Annual Conference of the Cognitive Science Society* (2015).

28. Kriegeskorte, N., Formisano, E., Sorger, B. & Goebel, R. Individual faces elicit distinct response patterns in human anterior temporal cortex. *Proceedings of the National Academy of Sciences* **104,** 20600–20605 (2007).

29. Anzellotti, S., Fairhall, S. L. & Caramazza, A. Decoding Representations of Face Identity That are Tolerant to Rotation. *Cerebral Cortex* **24,** 1988–1995 (2014).

30. Chang, L. & Tsao, D. Y. The Code for Facial Identity in the Primate Brain. *Cell* **169,** 1013–1028.e14 (2017).

31. Van Essen, D. C. *et al.* The Brain Analysis Library of Spatial maps and Atlases (BALSA) database. *NeuroImage* **144,** 270–274 (2017).

32. Griffiths, T. L., Chater, N., Kemp, C., Perfors, A. & Tenenbaum, J. B. Probabilistic models of cognition: exploring representations and inductive biases. *Trends in Cognitive Sciences* **14,** 357–364 (2010).

33. Ernst, M. O. & Banks, M. S. Humans integrate visual and haptic information in a statistically optimal fashion. *Nature* **415,** 429–433 (2002).

34. Wolpert, D. M. Probabilistic models in human sensorimotor control. *Human Movement Science* **26,** 511–524 (2007).

35. Weiss, Y., Simoncelli, E. P. & Adelson, E. H. Motion illusions as optimal percepts. *Nature Neuroscience* **5,** 598–604 (2002).

36. Körding, K. P. & Wolpert, D. M. Bayesian integration in sensorimotor learning. *Nature* **427,** 244–247 (2004).

37. MacKay, D. J. C. *Information theory, inference, and learning algorithms.* (Cambridge University Press, 2003).

38. Bishop, C. M. *Pattern recognition and machine learning.* (Springer, 2006).

39. Murphy, K. P. *Machine learning: a probabilistic perspective.* (MIT Press, 2012).

40. Dayan, P. & Abbott, L. F. *Theoretical neuroscience: computational and mathematical modeling of neural systems.* (Massachusetts Institute of Technology Press, 2001).

41. Abbott, L. F. Theoretical Neuroscience Rising. *Neuron* **60,** 489–495 (2008).

42. Olshausen, B. & Field, D. Sparse coding of sensory inputs. *Current Opinion in Neurobiology* **14,** 481–487 (2004).

43. Simoncelli, E. P. & Olshausen, B. A. Natural Image Statistics and Neural Representation. *Annual Review of Neuroscience* **24,** 1193–1216 (2001).

44. Carandini, M. & Heeger, D. J. Normalization as a canonical neural computation. *Nature Reviews Neuroscience* (2011). doi:10.1038/nrn3136

45. Chaudhuri, R. & Fiete, I. Computational principles of memory. *Nature Neuroscience* **19,** 394–403 (2016).

46. Shadlen, M. N. & Kiani, R. Decision Making as a Window on Cognition. *Neuron* **80,** 791–806 (2013).

47. Newsome, W. T., Britten, K. H. & Movshon, J. A. Neuronal correlates of a perceptual decision. *Nature* **341,** 52–54 (1989).

48. Wang, X.-J. Decision Making in Recurrent Neuronal Circuits. *Neuron* **60,** 215–234 (2008).

49. Diedrichsen, J., Shadmehr, R. & Ivry, R. B. The coordination of movement: optimal feedback control and beyond. *Trends in Cognitive Sciences* **14,** 31–39 (2010).

50. Kriegeskorte, N. Deep Neural Networks: A New Framework for Modeling Biological Vision and Brain Information Processing. *Annual Review of Vision Science* **1,** 417–446 (2015).

51. Yamins, D. L. K. & DiCarlo, J. J. Using goal-driven deep learning models to understand sensory cortex. *Nature Neuroscience* **19,** 356–365 (2016).

52. Krizhevsky, A., Sutskever, I. & Hinton, G. E. ImageNet Classification with Deep Convolutional Neural Networks. in *Advances in Neural Information Processing Systems 25* 1097–1105 (Curran Associates, Inc, 2012).

53. Silver, D. *et al.* Mastering the game of Go with deep neural networks and tree search. *Nature* **529,** 484–489 (2016).

54. Mnih, V. *et al.* Human-level control through deep reinforcement learning. *Nature* **518,** 529–533 (2015).

55. LeCun, Y., Bengio, Y. & Hinton, G. Deep learning. *Nature* **521,** 436–444 (2015).

56. Cohen, J. D. *et al.* Computational approaches to fMRI analysis. *Nature Neuroscience* **20,** 304–313 (2017).

57. Forstmann, B. U., Wagenmakers, E.-J., Eichele, T., Brown, S. & Serences, J. T. Reciprocal relations between cognitive neuroscience and formal cognitive models: opposites attract? *Trends in Cognitive Sciences* **15,** 272–279 (2011).

58. Deco, G., Tononi, G., Boly, M. & Kringelbach, M. L. Rethinking segregation and integration: contributions of whole-brain modelling. *Nature Reviews Neuroscience* **16,** 430–439 (2015).

59. Biswal, B., Zerrin Yetkin, F., Haughton, V. M. & Hyde, J. S. Functional connectivity in the motor cortex of resting human brain using echo-planar mri. *Magnetic Resonance in Medicine* **34,** 537–541 (1995).

60. A. Hyvarinen, J. K., E. Oja. *Independent Component Analysis.* (John Wiley & Sons, 2001).

61. Bassett, D. S. & Sporns, O. Network neuroscience. *Nature Neuroscience* **20,** 353–364 (2017).

62. Bullmore, E. T. & Bassett, D. S. Brain Graphs: Graphical Models of the Human Brain Connectome. *Annual Review of Clinical Psychology* **7,** 113–140 (2011).

63. Deco, G., Jirsa, V. K. & McIntosh, A. R. Emerging concepts for the dynamical organization of resting-state activity in the brain. *Nature Reviews Neuroscience* **12,** 43–56 (2011).

64. Friston, K. Dynamic causal modeling and Granger causality Comments on: The identification of interacting networks in the brain using fMRI: Model selection, causality and deconvolution. *NeuroImage* **58,** 303–305 (2011).

65. Roebroeck, A., Formisano, E. & Goebel, R. The identification of interacting networks in the brain using fMRI: Model selection, causality and deconvolution. *NeuroImage* **58,** 296–302 (2011).

66. Vicente, R., Wibral, M., Lindner, M. & Pipa, G. Transfer entropy—a model-free measure of effective connectivity for the neurosciences. *Journal of Computational Neuroscience* **30,** 45–67 (2011).

67. Anzellotti, S., Caramazza & Saxe, R. *Multivariate Pattern Connectivity*. (2016).

68. Bekinschtein, T. A. *et al.* Neural signature of the conscious processing of auditory regularities. *Proceedings of the National Academy of Sciences* **106,** 1672–1677 (2009).

69. Stephan, K. E. & Mathys, C. Computational approaches to psychiatry. *Current Opinion in Neurobiology* **25,** 85–92 (2014).

70. Tong, F. & Pratte, M. S. Decoding Patterns of Human Brain Activity. *Annual Review of Psychology* **63,** 483–509 (2012).

71. Haxby, J. V., Connolly, A. C. & Guntupalli, J. S. Decoding Neural Representational Spaces Using Multivariate Pattern Analysis. *Annual Review of Neuroscience* **37,** 435–456 (2014).

72. Dennett, D. C. *The intentional stance*. (MIT Press, 1987).

73. Kriegeskorte, N. Pattern-information analysis: From stimulus decoding to computational-model testing. *NeuroImage* **56,** 411–421 (2011).

74. Diedrichsen, J. & Kriegeskorte, N. Representational models: A common framework for understanding encoding, pattern-component, and representational-similarity analysis. *PLOS Computational Biology* **13,** e1005508 (2017).

75. Afraz, S.-R., Kiani, R. & Esteky, H. Microstimulation of inferotemporal cortex influences face categorization. *Nature* **442,** 692–695 (2006).

76. Parvizi, J. *et al.* Electrical Stimulation of Human Fusiform Face-Selective Regions Distorts Face Perception. *Journal of Neuroscience* **32,** 14915–14920 (2012).

77. *Spikes: exploring the neural code*. (MIT Press, 1997).

78. Norman, K. A., Polyn, S. M., Detre, G. J. & Haxby, J. V. Beyond mind-reading: multi-voxel pattern analysis of fMRI data. *Trends Cogn. Sci. (Regul. Ed.)* **10,** 424–430 (2006).

79. Kriegeskorte, N. & Kievit, R. A. Representational geometry: integrating cognition, computation, and the brain. *Trends in Cognitive Sciences* **17,** 401–412 (2013).

80. Haynes, J.-D. A Primer on Pattern-Based Approaches to fMRI: Principles, Pitfalls, and Perspectives. *Neuron* **87,** 257–270 (2015).

81. Jin, X. & Costa, R. M. Shaping action sequences in basal ganglia circuits. *Current Opinion in Neurobiology* **33,** 188–196 (2015).

82. DiCarlo, J. J. & Cox, D. D. Untangling invariant object recognition. *Trends in Cognitive Sciences* **11,** 333–341 (2007).

83. Kriegeskorte, N. *et al.* Matching Categorical Object Representations in Inferior Temporal Cortex of Man and Monkey. *Neuron* **60,** 1126–1141 (2008).

84. Douglas, P. K., Harris, S., Yuille, A. & Cohen, M. S. Performance comparison of machine learning algorithms and number of independent components used in fMRI decoding of belief vs. disbelief. *NeuroImage* **56,** 544–553 (2011).

85. Serences, J. T. & Saproo, S. Computational advances towards linking BOLD and behavior. *Neuropsychologia* **50,** 435–446 (2012).

86. Harrison, S. A. & Tong, F. Decoding reveals the contents of visual working memory in early visual areas. *Nature* **458,** 632–635 (2009).

87. Thirion, B. *et al.* Inverse retinotopy: Inferring the visual content of images from brain activation patterns. *NeuroImage* **33,** 1104–1116 (2006).

88. Miyawaki, Y. *et al.* Visual Image Reconstruction from Human Brain Activity using a Combination of Multiscale Local Image Decoders. *Neuron* **60,** 915–929 (2008).

89. Naselaris, T., Prenger, R. J., Kay, K. N., Oliver, M. & Gallant, J. L. Bayesian Reconstruction of Natural Images from Human Brain Activity. *Neuron* **63,** 902–915 (2009).

90. Naselaris, T. & Kay, K. N. Resolving Ambiguities of MVPA Using Explicit Models of Representation. *Trends in Cognitive Sciences* **19,** 551–554 (2015).

91. Mitchell, T. M. *et al.* Predicting Human Brain Activity Associated with the Meanings of Nouns. *Science* **320,** 1191–1195 (2008).

92. Kay, K. N., Naselaris, T., Prenger, R. J. & Gallant, J. L. Identifying natural images from human brain activity. *Nature* **452,** 352–355 (2008).

93. Dumoulin, S. O. & Wandell, B. A. Population receptive field estimates in human visual cortex. *NeuroImage* **39,** 647–660 (2008).

94. Diedrichsen, J., Ridgway, G. R., Friston, K. J. & Wiestler, T. Comparing the similarity and spatial structure of neural representations: A pattern-component model. *NeuroImage* **55,** 1665–1678 (2011).

95. Kriegeskorte, N. Representational similarity analysis – connecting the branches of systems neuroscience. *Frontiers in Systems Neuroscience* (2008). doi:10.3389/neuro.06.004.2008

96. Nili, H. *et al.* A Toolbox for Representational Similarity Analysis. *PLoS Computational Biology* **10,** e1003553 (2014).

97. Norman-Haignere, S., Kanwisher, N. G. & McDermott, J. H. Distinct Cortical Pathways for Music and Speech Revealed by Hypothesis-Free Voxel Decomposition. *Neuron* **88,** 1281–1296 (2015).

98. Devereux, B. J., Clarke, A., Marouchos, A. & Tyler, L. K. Representational Similarity Analysis Reveals Commonalities and Differences in the Semantic Processing of Words and Objects. *Journal of Neuroscience* **33,** 18906–18916 (2013).

99. Huth, A. G., de Heer, W. A., Griffiths, T. L., Theunissen, F. E. & Gallant, J. L. Natural speech reveals the semantic maps that tile human cerebral cortex. *Nature* **532,** 453–458 (2016).

100. Markram, H. The Blue Brain Project. *Nature Reviews Neuroscience* **7,** 153–160 (2006).

101. Eliasmith, C. & Trujillo, O. The use and abuse of large-scale brain models. *Current Opinion in Neurobiology* **25,** 1–6 (2014).

102. Eliasmith, C. *et al.* A Large-Scale Model of the Functioning Brain. *Science* **338,** 1202–1205 (2012).

103. Hassabis, D., Kumaran, D., Summerfield, C. & Botvinick, M. Neuroscience-Inspired Artificial Intelligence. *Neuron* **95,** 245–258 (2017).

104. Rumelhart, D. E., Hinton, G. E. & Williams, R. J. Learning representations by back-propagating errors. *Nature* **323,** 533–536 (1986).

105. Hornik, K. Approximation capabilities of multilayer feedforward networks. *Neural Networks* **4,** 251–257 (1991).

106. Goodfellow, I., Bengio, Y. & Courville, A. *Deep learning*. (The MIT Press, 2016).

107. Wyatte, D., Curran, T. & O'Reilly, R. The Limits of Feedforward Vision: Recurrent Processing Promotes Robust Object Recognition when Objects Are Degraded. *Journal of Cognitive Neuroscience* **24,** 2248–2261 (2012).

108. Spoerer, C. J., McClure, P. & Kriegeskorte, N. Recurrent Convolutional Neural Networks: A Better Model of Biological Object Recognition. *Frontiers in Psychology* **8,** (2017).

109. Hunt, L. T. & Hayden, B. Y. A distributed, hierarchical and recurrent framework for reward-based choice. *Nature Reviews Neuroscience* **18,** 172–182 (2017).

110. Schäfer, A. M. & Zimmermann, H.-G. Recurrent neural networks are universal approximators. *International Journal of Neural Systems* **17,** 253–263 (2007).

111. O'Reilly, R. C., Hazy, T. E., Mollick, J., Mackie, P. & Herd, S. Goal-Driven Cognition in the Brain: A Computational Framework. arxiv.org/abs/1404.7591 (2014).

112. Whittington, J. C. R. & Bogacz, R. An Approximation of the Error Backpropagation Algorithm in a Predictive Coding Network with Local Hebbian Synaptic Plasticity. *Neural Computation* **29,** 1229–1262 (2017).

113. Schiess, M., Urbanczik, R. & Senn, W. Somato-dendritic Synaptic Plasticity and Error-backpropagation in Active Dendrites. *PLOS Computational Biology* **12,** e1004638 (2016).

114. Marblestone, A. H., Wayne, G. & Kording, K. P. *Towards an integration of deep learning and neuroscience. Frontiers in computational neuroscience*, 10, 94 (2016).

115. Shadlen, M. N. & Shohamy, D. Decision Making and Sequential Sampling from Memory. *Neuron* **90,** 927–939 (2016).

116. Roelfsema, P. R. & Ooyen, A. van. Attention-Gated Reinforcement Learning of Internal Representations for Classification. *Neural Computation* **17,** 2176–2214 (2005).

117. Kingma, D. & Welling, M. Auto-encoding variational bayes. arxiv.org/abs/1312.6114 (2013).

118. Goodfellow, I. *et al.* Generative adversarial nets. *arXiv:1406.2661* (2014).

119. Yamins, D. L. K. *et al.* Performance-optimized hierarchical models predict neural responses in higher visual cortex. *Proceedings of the National Academy of Sciences* **111,** 8619–8624 (2014).

120. Khaligh-Razavi, S.-M. & Kriegeskorte, N. Deep Supervised, but Not Unsupervised, Models May Explain IT Cortical Representation. *PLoS Computational Biology* **10,** e1003915 (2014).

121. Cadieu, C. F. *et al.* Deep Neural Networks Rival the Representation of Primate IT Cortex for Core Visual Object Recognition. *PLoS Computational Biology* **10,** e1003963 (2014).

122. Guclu, U. & van Gerven, M. A. J. Deep Neural Networks Reveal a Gradient in the Complexity of Neural Representations across the Ventral Stream. *Journal of Neuroscience* **35,** 10005–10014 (2015).

123. Eickenberg, M., Gramfort, A., Varoquaux, G. & Thirion, B. Seeing it all: Convolutional network layers map the function of the human visual system. *NeuroImage* (2016). doi:10.1016/j.neuroimage.2016.10.001

124. Cichy, R. M., Khosla, A., Pantazis, D., Torralba, A. & Oliva, A. Comparison of deep neural networks to spatio-temporal cortical dynamics of human visual object recognition reveals hierarchical correspondence. *Scientific Reports* **6,** (2016).

125. Hong, H., Yamins, D. L. K., Majaj, N. J. & DiCarlo, J. J. Explicit information for category-orthogonal object properties increases along the ventral stream. *Nature Neuroscience* **19,** 613–622 (2016).

126. Kubilius, J., Bracci, S. & Op de Beeck, H. P. Deep Neural Networks as a Computational Model for Human Shape Sensitivity. *PLOS Computational Biology* **12,** e1004896 (2016).

127. Jozwik, K. M., Kriegeskorte, N., Storrs, K. R. & Mur, M. Deep Convolutional Neural Networks Outperform Feature-Based But Not Categorical Models in Explaining Object Similarity Judgments. *Frontiers in Psychology* **8,** (2017).

128. Sutton, R. & Barto, A. *Reinforcement Learning: An Introduction.* (Cambridge MIT Press, 1998).

129. Dayan, P. & Daw, N. D. Decision theory, reinforcement learning, and the brain. *Cognitive, Affective, & Behavioral Neuroscience* **8,** 429–453 (2008).

130. Silver, D. *et al.* Mastering the game of Go without human knowledge. *Nature* **550,** 354–359 (2017).

131. Serre, T., Wolf, L., Bileschi, S., Riesenhuber, M. & Poggio, T. Robust Object Recognition with Cortex-Like Mechanisms. *IEEE Transactions on Pattern Analysis and Machine Intelligence* **29,** 411–426 (2007).

132. Cox, D. D. & Dean, T. Neural Networks and Neuroscience-Inspired Computer Vision. *Current Biology* **24,** R921–R929 (2014).

133. Moore, C. & Mertens, S. *The nature of computation.* (Oxford University Press, 2011).

134. Borst, J. & Taatgen & Anderson, J. *Using the ACT-R Cognitive Architecture in combination with fMRI data. In B. U. Forstmann, & E.-J. Wagenmakers (Eds.), An Introduction to Model-Based Cognitive Neuroscience.* (Springer: New York, 2014).

135. Daw, N. D. & Dayan, P. The algorithmic anatomy of model-based evaluation. *Philosophical Transactions of the Royal Society B: Biological Sciences* **369,** 20130478–20130478 (2014).

136. Lengyel, M. & Dayan, P. Hippocampal Contributions to Control: The Third Way. *NIPS* 889–896 (2008).

137. Gershman, S. J. & Daw, N. D. Reinforcement Learning and Episodic Memory in Humans and Animals: An Integrative Framework. *Annual Review of Psychology* **68,** 101–128 (2017).

138. O'Doherty, J. P., Cockburn, J. & Pauli, W. M. Learning, Reward, and Decision Making. *Annual Review of Psychology* **68,** 73–100 (2017).

139. Schultz, W., Dayan, P. & Montague, P. R. A neural substrate of prediction and reward. *Science* **275,** 1593–1599 (1997).

140. Sutton, R. Integrated architectures for learning, planning, and reacting based on approximating dynamic programming. *Proceedings of the seventh international conference on machine learning* 216–224 (1990).

141. Dayan, P. Improving Generalization for Temporal Difference Learning: The Successor Representation. *Neural Computation* **5,** 613–624 (1993).

142. Daw, N. D., Niv, Y. & Dayan, P. Uncertainty-based competition between prefrontal and dorsolateral striatal systems for behavioral control. *Nature Neuroscience* **8,** 1704–1711 (2005).

143. Ma, W. J. Organizing probabilistic models of perception. *Trends in Cognitive Sciences* **16,** 511–518 (2012).

144. Fiser, J., Berkes, P., Orbán, G. & Lengyel, M. Statistically optimal perception and learning: from behavior to neural representations. *Trends in Cognitive Sciences* **14,** 119–130 (2010).

145. Tenenbaum, J. B., Kemp, C., Griffiths, T. L. & Goodman, N. D. How to Grow a Mind: Statistics, Structure, and Abstraction. *Science* **331,** 1279–1285 (2011).

146. Yuille, A. & Kersten, D. Vision as Bayesian inference: analysis by synthesis? *Trends in Cognitive Sciences* **10,** 301–308 (2006).

147. Helmholtz, H. *Handbuch der Physiologischen Optik*. (Dover. New York, 1860).

148. Lasserre, J. A., Bishop, C. M. & Minka, T. P. Principled Hybrids of Generative and Discriminative Models. in **1,** 87–94 (IEEE, 2006).

149. Gershman, S. J., Horvitz, E. J. & Tenenbaum, J. B. Computational rationality: A converging paradigm for intelligence in brains, minds, and machines. *Science* **349,** 273–278 (2015).

150. Simon, H. A. Bounded Rationality. in *Utility and Probability* (eds. Eatwell, J., Milgate, M. & Newman, P.) 15–18 (Palgrave Macmillan UK, 1990). doi:10.1007/978-1-349-20568-4_5

151. Griffiths, T. L., Lieder, F. & Goodman, N. D. Rational Use of Cognitive Resources: Levels of Analysis Between the Computational and the Algorithmic. *Topics in Cognitive Science* **7,** 217–229 (2015).

152. Srikumar, V., Kundu, G. & Roth, D. On Amortizing Inference Cost for Structured Prediction. *Proceeedings of the 2012 Joint Conference on Empirical Methods in Natural Language Processing and Computational Natural Language Learning* 1114–1124 (2012).

153. Bengio, Y., Scellier, B., Bilaniuk, O., Sacramento, J. & Senn, W. Feedforward initialization for fast inference of deep generative networks is biologically plausible. *arXiv:1606.01651* (2016).

154. Tversky, A. & Kahneman, D. Judgment under Uncertainty: Heuristics and Biases. in *Utility, Probability, and Human Decision Making* (eds. Wendt, D. & Vlek, C.) 141–162 (Springer Netherlands, 1975). doi:10.1007/978-94-010-1834-0_8

155. Ghahramani, Z. Bayesian non-parametrics and the probabilistic approach to modelling. *Philosophical Transactions of the Royal Society A: Mathematical, Physical and Engineering Sciences* **371,** 20110553–20110553 (2012).

156. Lake, B. M., Salakhutdinov, R. & Tenenbaum, J. B. Human-level concept learning through probabilistic program induction. *Science* **350,** 1332–1338 (2015).

157. Ullman, T. D., Spelke, E., Battaglia, P. & Tenenbaum, J. B. Mind Games: Game Engines as an Architecture for Intuitive Physics. *Trends in Cognitive Sciences* **21,** 649–665 (2017).

158. Battaglia, P. W., Hamrick, J. B. & Tenenbaum, J. B. Simulation as an engine of physical scene understanding. *Proceedings of the National Academy of Sciences* **110,** 18327–18332 (2013).

159. Kubricht, J. R., Holyoak, K. J. & Lu, H. Intuitive Physics: Current Research and Controversies. *Trends in Cognitive Sciences* **21,** 749–759 (2017).

160. Pantelis, P. C. *et al.* Inferring the intentional states of autonomous virtual agents. *Cognition* **130,** 360–379 (2014).

161. Pouget, A., Beck, J. M., Ma, W. J. & Latham, P. E. Probabilistic brains: knowns and unknowns. *Nature Neuroscience* **16,** 1170–1178 (2013).

162. Orhan, A. E. & Ma, W. J. Efficient probabilistic inference in generic neural networks trained with non-probabilistic feedback. *Nature Communications* **8,** (2017).

163. Tervo, D. G. R., Tenenbaum, J. B. & Gershman, S. J. Toward the neural implementation of structure learning. *Current Opinion in Neurobiology* **37,** 99–105 (2016).

164. Berkes, P., Orban, G., Lengyel, M. & Fiser, J. Spontaneous Cortical Activity Reveals Hallmarks of an Optimal Internal Model of the Environment. *Science* **331,** 83–87 (2011).

165. Buesing, L., Bill, J., Nessler, B. & Maass, W. Neural Dynamics as Sampling: A Model for Stochastic Computation in Recurrent Networks of Spiking Neurons. *PLoS Computational Biology* **7,** e1002211 (2011).

166. Haefner, R. M., Berkes, P. & Fiser, J. Perceptual Decision-Making as Probabilistic Inference by Neural Sampling. *Neuron* **90,** 649–660 (2016).

167. Aitchison, L. & Lengyel, M. The Hamiltonian Brain: Efficient Probabilistic Inference with Excitatory-Inhibitory Neural Circuit Dynamics. *PLOS Computational Biology* (2016).

168. Sanborn, A. N. & Chater, N. Bayesian Brains without Probabilities. *Trends in Cognitive Sciences* **20,** 883–893 (2016).

169. Dasgupta, I., Schulz, E., Goodman, N. & Gershman, S. Amortized Hypothesis Generation. (2017). doi:10.1101/137190

170. Rezende, D., Mohamed, S., Danihelka, I., Gregor, K. & Wierstra, D. One-shot generalization in deep generative models. *Proceedings of the 33rd International Conference on International Conference on Machine Learning* **48,** 1521–1529 (2016).

171. Daw, N. D., O'Doherty, J. P., Dayan, P., Seymour, B. & Dolan, R. J. Cortical substrates for exploratory decisions in humans. *Nature* **441,** 876–879 (2006).

172. Krakauer, J. W., Ghazanfar, A. A., Gomez-Marin, A., MacIver, M. A. & Poeppel, D. Neuroscience Needs Behavior: Correcting a Reductionist Bias. *Neuron* **93,** 480–490 (2017).

173. Gomez-Marin, A., Paton, J. J., Kampff, A. R., Costa, R. M. & Mainen, Z. F. Big behavioral data: psychology, ethology and the foundations of neuroscience. *Nature Neuroscience* **17,** 1455–1462 (2014).

174. Marr, D. *Vision: a computational investigation into the human representation and processing of visual information*. (MIT Press, 2010).

175. Gal, Y. & Ghahramani, Z. Dropout as a Bayesian Approximation: Representing Model Uncertainty in Deep Learning. *arXiv:1506.02142* (2016).

176. Wu, J., Yildirim, I., Lim, J., Freeman, B. & Tenenbaum, J. Galileo: Perceiving physical object properties by integrating a physics engine with deep learning. *Annual conference of the cognitive science society* 127–135 (2015).

177. Kulkarni, T., Whitney, W., Kohli, P. & Tenebaum, J. Deep convolutional inverse graphics network. *Advances in Neural Processing Systems* 2539–2547 (2015).

178. Dayan, P., Hinton, G. E., Neal, R. M. & Zemel, R. S. The Helmholtz Machine. *Neural Computation* **7,** 889–904 (1995).

179. Hinton, G., Sejnowski, T. & Ackley, D. *Boltzmann machines: Constraint satisfaction networks that learn*. (Carnegie-Mellon University, 1984).

180. Hinton, G. E. Reducing the Dimensionality of Data with Neural Networks. *Science* **313,** 504–507 (2006).

181. Silver, D. *et al.* Mastering the game of Go with deep neural networks and tree search. *Nature* **529,** 484–489 (2016).

182. Hochreiter, S. & Schmidhuber, J. Long Short-Term Memory. *Neural Computation* **9,** 1735–1780 (1997).

183. Graves, A. *et al.* Hybrid computing using a neural network with dynamic external memory. *Nature* **538,** 471–476 (2016).

184. Kumaran, D., Hassabis, D. & McClelland, J. L. What Learning Systems do Intelligent Agents Need? Complementary Learning Systems Theory Updated. *Trends in Cognitive Sciences* **20,** 512–534 (2016).

185. Kandel, E.R., Schwartz J.H., Jessell T.M., Siegelbaum S.A., Hudspeth A.J. *Principles of Neural Science*. (McGraw-Hill, 2013).

186. Bastos, A. M. *et al.* Canonical Microcircuits for Predictive Coding. *Neuron* **76,** 695–711 (2012).

187. Larkum, M. A cellular mechanism for cortical associations: an organizing principle for the cerebral cortex. *Trends in Neurosciences* **36,** 141–151 (2013).

188. Fries, P. A mechanism for cognitive dynamics: neuronal communication through neuronal coherence. *Trends in Cognitive Sciences* **9,** 474–480 (2005).

189. Jia Deng *et al.* ImageNet: A large-scale hierarchical image database. in 248–255 (IEEE, 2009). doi:10.1109/CVPR.2009.5206848

190. Griffiths, T. L. Manifesto for a new (computational) cognitive revolution. *Cognition* **135,** 21–23 (2015).

191. Ahrens, M. B. *et al.* Brain-wide neuronal dynamics during motor adaptation in zebrafish. *Nature* (2012). doi:10.1038/nature11057

192. Kietzmann, T., McClure, P. & Kriegeskorte, N. Deep Neural Networks In Computational Neuroscience. *bioRxiv, 133504.* (2017).

193. Beattie, C. *et al.* Deepmind lab. *arXiv:1612.03801* (2016).

194. Naselaris, T., Bassett, D.S., Fletcher, A.K., Kording, K., Kriegeskorte, N., Nienborg, H., Poldrack, R.A., Shohamy, D. and Kay, K.et al. Cognitive Computational Neuroscience: a new conference for an emerging discipline. *Trends Cognitive Sciences* **22**, 365–367 (2018).